\documentclass[aps,prl,twocolumn,showpacs,superscriptaddress,groupedaddress]{revtex4}  
\usepackage{graphicx}  
\usepackage{dcolumn}   
\usepackage{bm}        
\usepackage{amssymb}   
\usepackage{amsmath}
\usepackage{color}
\begin{document}


\title{Demonstration of an effective ultrastrong coupling between two oscillators}

\author{D. Markovi\'c}
\affiliation{Laboratoire Pierre Aigrain, D\'epartement de physique de l'ENS, \'Ecole normale sup\'erieure, PSL Research University, Universit\'e Paris Diderot, Sorbonne Paris Cit\'e, Sorbonne Universit\'es, UPMC Univ. Paris 06, CNRS,
75005 Paris, France
}

\author{S. Jezouin}
\affiliation{Laboratoire Pierre Aigrain, D\'epartement de physique de l'ENS, \'Ecole normale sup\'erieure, PSL Research University, Universit\'e Paris Diderot, Sorbonne Paris Cit\'e, Sorbonne Universit\'es, UPMC Univ. Paris 06, CNRS,
75005 Paris, France
}

\author{Q. Ficheux}

\affiliation{Universit\'e Lyon, ENS de Lyon, Universit\'e Claude Bernard Lyon 1, CNRS, Laboratoire de Physique, F-69342
Lyon, France}
\affiliation{Laboratoire Pierre Aigrain, D\'epartement de physique de l'ENS, \'Ecole normale sup\'erieure, PSL Research University, Universit\'e Paris Diderot, Sorbonne Paris Cit\'e, Sorbonne Universit\'es, UPMC Univ. Paris 06, CNRS,
75005 Paris, France
}

\author{S. Fedortchenko}
\affiliation{Laboratoire Mat\' eriaux et Ph\' enom\`enes Quantiques, Sorbonne Paris Cit\' e, Universit\' e Paris Diderot, CNRS UMR 7162, 75013, Paris, France}

\author{S. Felicetti}
\affiliation{Laboratoire Mat\' eriaux et Ph\' enom\`enes Quantiques, Sorbonne Paris Cit\' e, Universit\' e Paris Diderot, CNRS UMR 7162, 75013, Paris, France}

\author{T. Coudreau}
\affiliation{Laboratoire Mat\' eriaux et Ph\' enom\`enes Quantiques, Sorbonne Paris Cit\' e, Universit\' e Paris Diderot, CNRS UMR 7162, 75013, Paris, France}

\author{P. Milman}
\affiliation{Laboratoire Mat\' eriaux et Ph\' enom\`enes Quantiques, Sorbonne Paris Cit\' e, Universit\' e Paris Diderot, CNRS UMR 7162, 75013, Paris, France}

\author{Z. Leghtas}
\affiliation{Centre Automatique et Syst\`emes, Mines ParisTech, PSL Research University,
60 Boulevard Saint-Michel, 75272 Paris Cedex 6, France.}
\affiliation{Laboratoire Pierre Aigrain, D\'epartement de physique de l'ENS, \'Ecole normale sup\'erieure, PSL Research University, Universit\'e Paris Diderot, Sorbonne Paris Cit\'e, Sorbonne Universit\'es, UPMC Univ. Paris 06, CNRS,
75005 Paris, France
}
\affiliation{QUANTIC team, INRIA de Paris, 2 Rue Simone Iff, 75012 Paris, France}

\author{B. Huard}
\affiliation{Universit\'e Lyon, ENS de Lyon, Universit\'e Claude Bernard Lyon 1, CNRS, Laboratoire de Physique, F-69342
Lyon, France}
\affiliation{Laboratoire Pierre Aigrain, D\'epartement de physique de l'ENS, \'Ecole normale sup\'erieure, PSL Research University, Universit\'e Paris Diderot, Sorbonne Paris Cit\'e, Sorbonne Universit\'es, UPMC Univ. Paris 06, CNRS,
75005 Paris, France
}

\date{\today}

\begin{abstract}
When the coupling rate between two quantum systems becomes as large as their characteristic frequencies, it induces dramatic effects on their dynamics and even on the nature of their ground state. The case of a qubit coupled to a harmonic oscillator in this ultrastrong coupling regime has been investigated theoretically and experimentally. Here, we explore the case of two harmonic oscillators in the ultrastrong coupling regime. Specifically, we realize an analog quantum simulation of this coupled system by dual frequency pumping a nonlinear superconducting circuit. The pump amplitudes directly tune the effective coupling rate. We observe spectroscopic signature of a mode hybridization that is characteristic of the ultrastrong coupling. Further we experimentally demonstrate a key property of the ground state of this simulated ultrastrong coupling between modes by observing simultaneous single-mode and two-mode squeezing of the radiated field below vacuum fluctuations.
\end{abstract}

\maketitle

The ultrastrong coupling regime characterizes quantum systems that are coupled at a rate so large that it reaches a significant fraction of their characteristic frequencies. Not only do the systems hybridize into a joint entity, but their joint dynamics cannot be captured by common approximations that are valid at lower coupling rates such as the rotating wave approximation. Beyond the unusual spectrum it produces~\cite{Braak2011}, the ultrastrong coupling modifies the ground state of the coupled systems in such a way that the systems get entangled at zero temperature and carry excitations in the basis of the isolated systems.
Releasing these excitations so that they can be detected or used as a source of work requires to abruptly switch off or modulate in time the ultrastrong interaction~\cite{Ciuti2006, DeLiberato2007}, which is still out of experimental reach. Alternatively, one can gain an insight on the properties of the ground state of an ultrastrongly coupled system by performing quantum simulations~\cite{Ballester2012}. Besides its fundamental interest the ultrastrong coupling has interesting applications for quantum computing  such as ultrafast two-qubit gates~\cite{Romero2012}, quantum memories~\cite{Kyaw2015,Stassi2018} or photon transfer through cavity arrays~\cite{Baust2016,PhysRevA.89.013853}. It has motivated experiments in various systems including cavity polaritons~\cite{Günter2009}, superconducting circuits~\cite{Niemczyk2010,Baust2016,Yoshihara2016,Forn-Diaz2016,Forn-Diaz2017,Bosman2017a,Langford2017,Braumuller2017}, cavity magnons~\cite{Goryachev2014} and bidimensional electron gases in THz cavities~\cite{Scalari2012, Zhang2016a}. 

\begin{figure}
\includegraphics[scale=0.45]{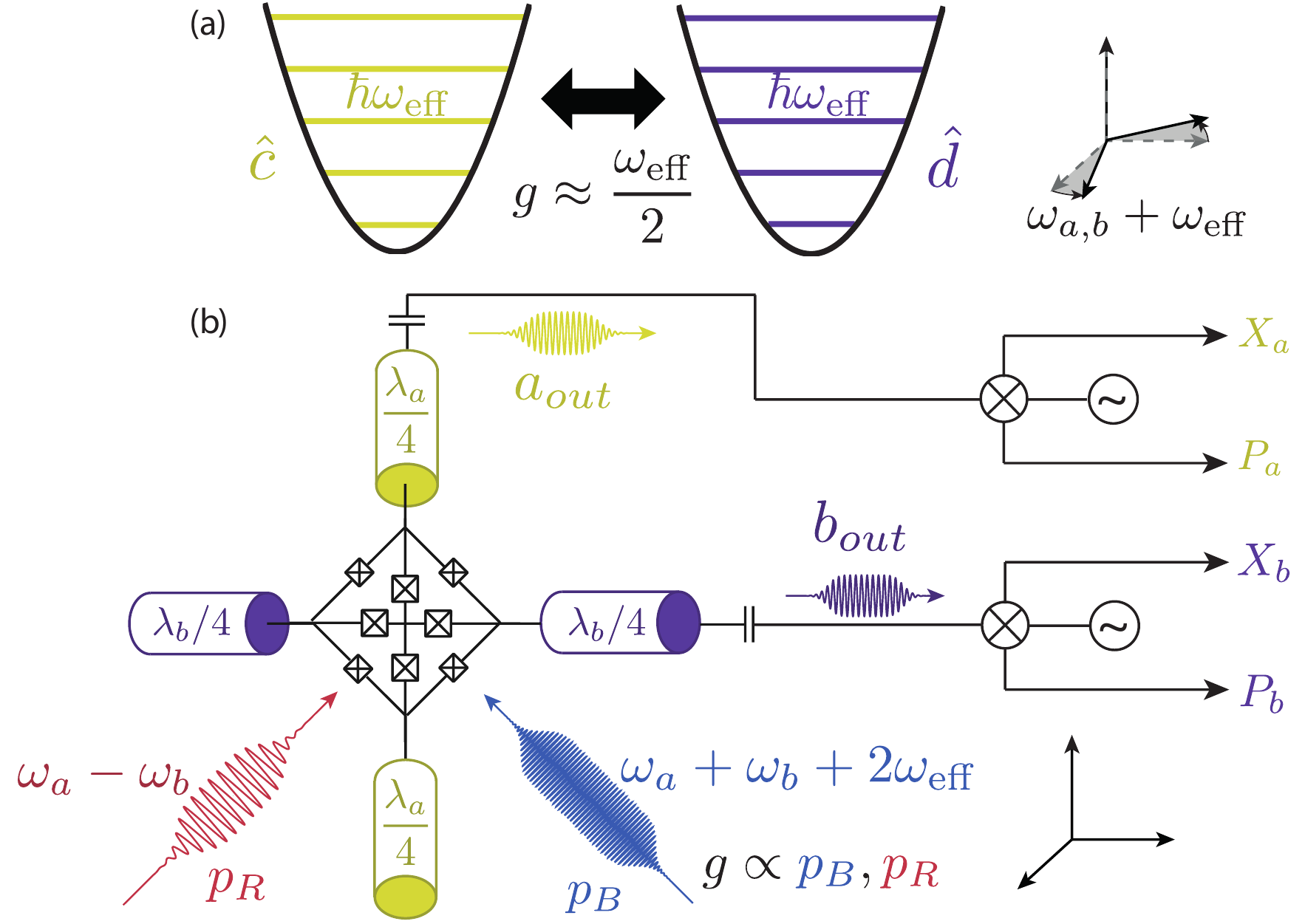}
\caption{(a) We perform an analog quantum simulation of a system made of two degenerate harmonic oscillators whose resonance frequency is $\omega_{\textrm{eff}}$ and that are ultrastrongly coupled with a rate $g$ of the order of $\omega_{\textrm{eff}}/2$.  (b) Multi-driven Josephson mixer implementing the ultrastrongly coupled system in (a) in the adequate rotating frame. The Josephson mixer consists of two microwave resonators resonating at frequencies $\omega_a$ and $\omega_b$, which are coupled via a Josephson Ring modulator~\cite{Bergeal2010,Roch2012}. By simultaneously pumping it with a ``red'' tone at $\omega_{p_R} = \omega_a - \omega_b$ and a ``blue'' tone at $\omega_{p_B} = \omega_a + \omega_b + 2 \omega_{\textrm{eff}}$, two effective modes at frequencies $\omega_{\textrm{eff}}$ are coupled at a rate $g$ that is proportional to the tunable pump amplitudes $p_B$ and $p_R$.}
\label{schema}
\end{figure}

In the case of a two level system ultrastrongly coupled to a harmonic oscillator -- the quantum Rabi model -- both digital~\cite{Langford2017} and analog~\cite{Braumuller2017} quantum simulations  have recently been performed on superconducting qubits to probe the characteristics of the ground state. Interestingly, this regime can also be reached between two coupled harmonic oscillators and requires experimental investigation. In this letter, we use an analog approach to mimic two harmonic oscillators in the ultrastrong coupling regime~\cite{Fedortchenko2017}. Specifically, we realize a multi-driven superconducting circuit that behaves, in a rotating frame, as two degenerate coupled harmonic oscillators resonating at frequency $\omega_{\textrm{eff}}$ (see Fig.~\ref{schema}). This analog quantum simulation allows us to map properties of the ground state onto the output signals of the circuit, making possible to observe fundamental features of the ultrastrong coupling that are not accessible otherwise.

The superconducting circuit is a Josephson mixer~\cite{Bergeal2010, Bergeal2010b,Roch2012}, which couples two microwave modes $a$ and $b$ through a three-wave mixing interaction involving a stiff pump~\cite{Bergeal2010b} signal applied with a complex amplitude $p$ and frequency $\omega_p$ (see Fig.~\ref{schema}b). It is described by the Hamiltonian
\begin{equation}
\hat{H} = \hbar \omega_a \hat{a}^\dagger \hat{a} + \hbar \omega_b \hat{b}^\dagger \hat{b} + \hbar \chi (p + p^\ast)(\hat{a}+\hat{a}^\dagger)(\hat{b}+\hat{b}^\dagger). \label{H_JPC}
\end{equation}Here, the fundamental modes resonate at frequencies $\omega_a = 2 \pi \times 8.477$~GHz and $\omega_b = 2 \pi \times 6.476$~GHz, and they are coupled to independent transmission lines at rates $\kappa_a = 2 \pi \times (19\pm 1)$~MHz and $\kappa_b = 2 \pi \times (22\pm 1)$~MHz. In our circuits, the parametric coupling rate $2\chi |p|$ is always much smaller than the frequencies $\omega_{a,b}$ so that the ultrastrong coupling regime cannot be reached in the lab frame. However, we have shown in Ref.~\cite{Fedortchenko2017} that by adequately pumping the circuit, an effective system in ultrastrong coupling regime emerges. 

We place ourselves in the reference frame that rotates at frequency $\omega_{a} + \omega_{\textrm{eff}}$ for mode $a$ and frequency $\omega_{b} + \omega_{\textrm{eff}}$ for mode $b$~\cite{Wustmann2017}. Here, the frequency $\omega_{\textrm{eff}}$ will be the degenerate resonant frequency of each effective system and is arbitrarily chosen. We thus define two effective modes whose canonical operators are $\hat{c}=e^{-i(\omega_a+\omega_\mathrm{eff})t}\hat{a}$ and $\hat{d}=e^{-i(\omega_b+\omega_\mathrm{eff})t}\hat{b}$. Their coupling term can be engineered into the desired form by applying a pump, which is the sum of two tones referred to as the ``red'' and ``blue'' pumps. The blue pump at frequency $\omega_{p_B} = \omega_a + \omega_b + 2\omega_{\textrm{eff}}$ has an amplitude $p_B$ and the red pump at $\omega_{p_R}= \omega_a - \omega_b$ has an amplitude $p_R$. In the rotating wave approximation (valid for $a$ and $b$ modes), the three-wave mixing interaction reduces to a sum of two terms: a parametric down-conversion term $\hat{H}_{pdc} = \hbar g_B (\hat{c}^\dagger \hat{d}^\dagger + \hat{c} \hat{d})$, where $g_B = \chi |p_B|$ is the coupling rate of the ``blue'' pump, and a parametric frequency conversion term $\hat{H}_{conv} = \hbar g_R (\hat{c}^\dagger \hat{d} + \hat{c} \hat{d}^\dagger) $, where $g_R = \chi |p_R|$ is the coupling rate of the ``red'' pump. Finally, by simultaneously applying the two pumps with tuned amplitudes $p_B$ and $p_R$ such that $g_B = g_R\equiv g$, we obtain the effective Hamiltonian
\begin{equation}
\hat{H}_{\textrm{eff}} = -\hbar \omega_{\textrm{eff}}\hat{c}^\dagger \hat{c} - \hbar \omega_{\textrm{eff}}\hat{d}^\dagger \hat{d} + \hbar g (\hat{c}+\hat{c}^\dagger)(\hat{d}+ \hat{d}^\dagger).
\label{USC_H}
\end{equation}
It reaches ultrastrong coupling when $g$ is of the order of $\omega_{\mathrm{eff}}/10$ or greater.

We first characterize the system by measuring the power spectral density of the emitted radiation in various pumping configurations. When only the ``red'' pump is applied, it is possible to check that the system is close to its quantum ground state at rest. This pumping scheme corresponds to the conversion of photons from $a$ to $b$ modes~\cite{Abdo2013a,Flurin2015} for a range of amplitudes $g_R$. Therefore the difference between the output spectral power $S_a^{ON}$ of the $a$ mode when the pump is on and the output spectral power $S_a^{OFF}$ when the pump is off is proportional to $\hbar\omega_a\coth\left(\hbar\omega_b/2k_BT\right)-\hbar\omega_a\coth\left(\hbar\omega_a/2k_BT\right)$. The fact that we could not observe any change in the output spectral density compared to the case where the pump is off (not shown) thus indicates that $k_BT\ll \hbar\omega_{a,b}$ at rest.

\begin{figure}[h!]
\includegraphics[width=8.5cm]{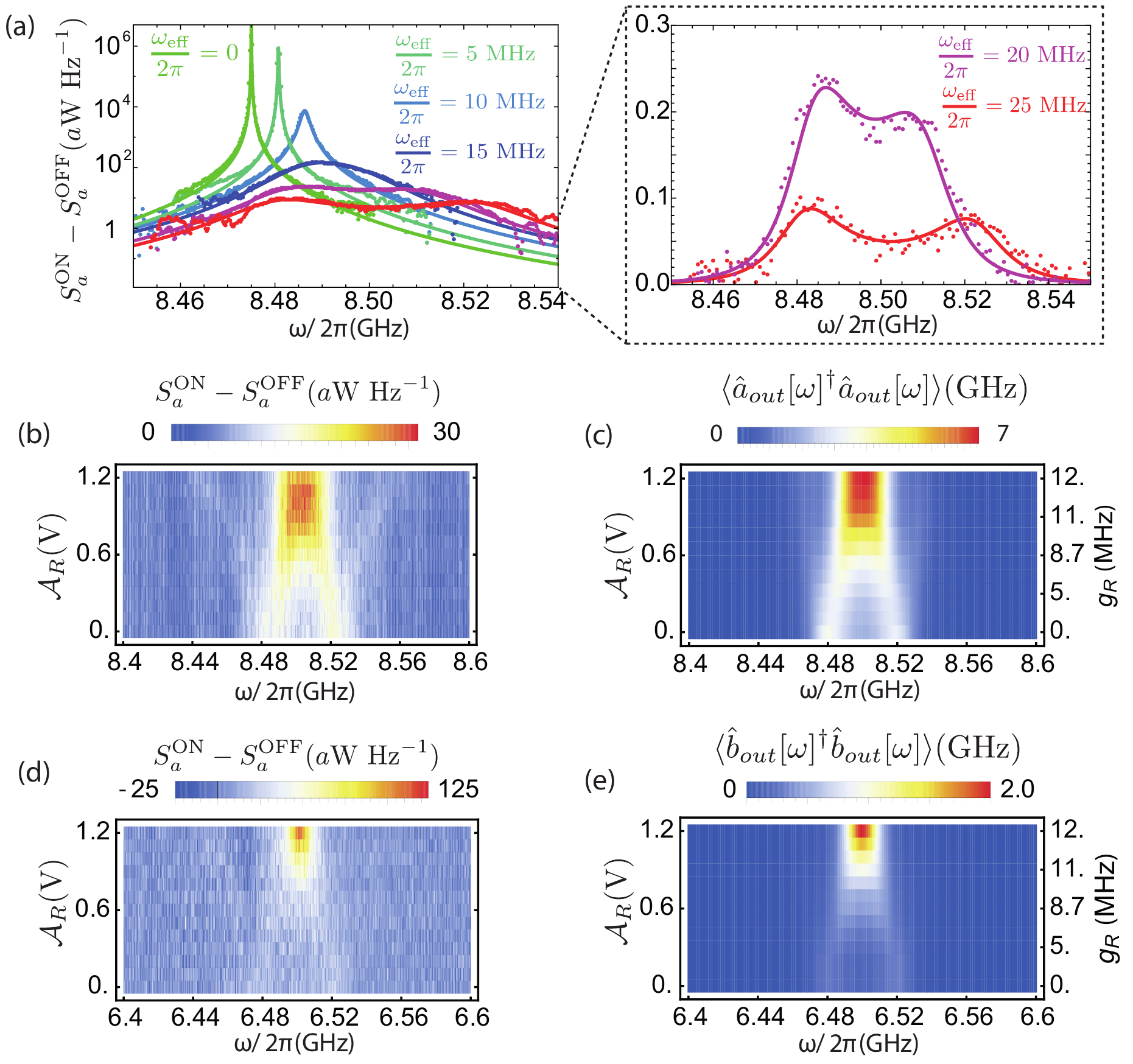}
\caption{(a) Measured power spectral density of the radiation emitted by the mode $a$ as a function of frequency, for various color encoded detunings $\omega_{\textrm{eff}}$ of the blue pump. Full lines correspond to the average number of emitted photons calculated, up to an overall factor, using quantum Langevin equation formalism and the input-output theory. All spectra are measured with the same pump power (corresponding to $g_B=2\pi\times 11.5~\mathrm{MHz}$) and are represented minus an offset corresponding to the spectrum that is measured when the pump is off ($S^\mathrm{ON}-S^\mathrm{OFF}$). (b) Measured power spectral density of the radiation emitted from the mode $a$ as a function of probe frequency $\omega$ and amplitude $\mathcal{A}_R$ of the ``red'' pump (voltage on the mixer that generates the pump). The ``blue'' pump amplitude $\mathcal{A}_B$ is fixed such that the coupling $g_B = 2 \pi \times 12.6$~MHz and its detuning is set to $\omega_{\textrm{eff}} = 2 \pi \times 26$~MHz. (c) Average photon emission rate out of the $a$ mode as a function of $\omega$ and $g_R$ calculated using quantum Langevin equation formalism and the input-output theory. (d,e) Similar plots in the case of the $b$ mode.}
\label{spectro}
\end{figure}

Now, let us consider the case when only the ``blue'' pump is applied. When the pump frequency is the sum $\omega_p = \omega_a + \omega_b$, the Josephson mixer acts as a parametric amplifier~\cite{Bergeal2010}. If the input ports are undriven and thus in the vacuum state, the amplification of vacuum fluctuations generates spatially separated propagating two-mode squeezed states (Einstein Podolsky Rosen or EPR states)~\cite{Flurin2012a}. For the present experiment instead, the pump frequency needs to be detuned by $2\omega_{\textrm{eff}}$, which leads to unexplored consequences. The impact of this detuning on the measured power spectral density at the output of port $a$ can be seen in Fig.~\ref{spectro}a for various values of $\omega_{\textrm{eff}}$ and a single pump power. For $\omega_{\textrm{eff}}<0$ (not shown), the device is in the regime of a strongly non-linear response detrimental to squeezing~\cite{Wustmann2017}. For the scope of the present work, we thus set $\omega_{\textrm{eff}} \geq 0$. As the detuning increases, the power spectral density decreases and three regimes can be identified. For the smallest detuning $\omega_{\textrm{eff}}/2\pi\lesssim 5~\mathrm{MHz}$, the chosen pump power is large enough for the Josephson mixer to enter the parametric oscillation regime and a single spectral peak develops at a frequency close to $\omega_a+\omega_\textrm{eff}$. In practice it is slightly shifted by Kerr effect. A broad single peak corresponding to the regime of parametric amplification of vacuum fluctuations can be observed when the detuning $10~\mathrm{MHz}\lesssim\omega_{\textrm{eff}}/2\pi\lesssim 15~\mathrm{MHz}$ is still smaller than the resonator bandwidth. Expectedly, this regime could also be observed for zero detuning but with smaller pump power. For even larger detuning $\omega_{\textrm{eff}}\geq \kappa_a,\kappa_b\approx2\pi\times 20~\mathrm{MHz}$, two peaks are resolved at frequencies $\omega_{a}$ and $\omega_{a} + 2 \omega_{\textrm{eff}}$ (see right panel of Fig.~\ref{spectro}a). Reciprocally, two peaks at $\omega_{b}$ and $\omega_{b} + 2 \omega_{\textrm{eff}}$ can be observed in the spectral density on the output of the $b$ mode. These peaks can be simply understood by realizing that the three wave mixing term of Eq.~(\ref{H_JPC}) consists, for any value of $\widetilde{\omega}$, in converting a pump photon at frequency $\omega_a+\omega_b+2\omega_{\mathrm{eff}}$ into a photon at frequency $\omega_a+\omega_{\mathrm{eff}}+\widetilde{\omega}$ on mode $a$ and a photon at frequency $\omega_b+\omega_{\mathrm{eff}}-\widetilde{\omega}$ on mode $b$. Therefore, the frequency of these photons is at the resonance of the $a$ mode for $\widetilde{\omega}=-\omega_{\mathrm{eff}}$ and of the $b$ mode for $\widetilde{\omega}=\omega_{\mathrm{eff}}$. The two peaks on the spectral power density of the $a$ mode (Fig.~\ref{spectro}a) thus correspond to the resonance of $a$ for the peak at $\omega_a$ and of $b$ for the peak at $\omega_a+2\omega_\mathrm{eff}$.
\begin{figure}
\includegraphics[scale=0.45]{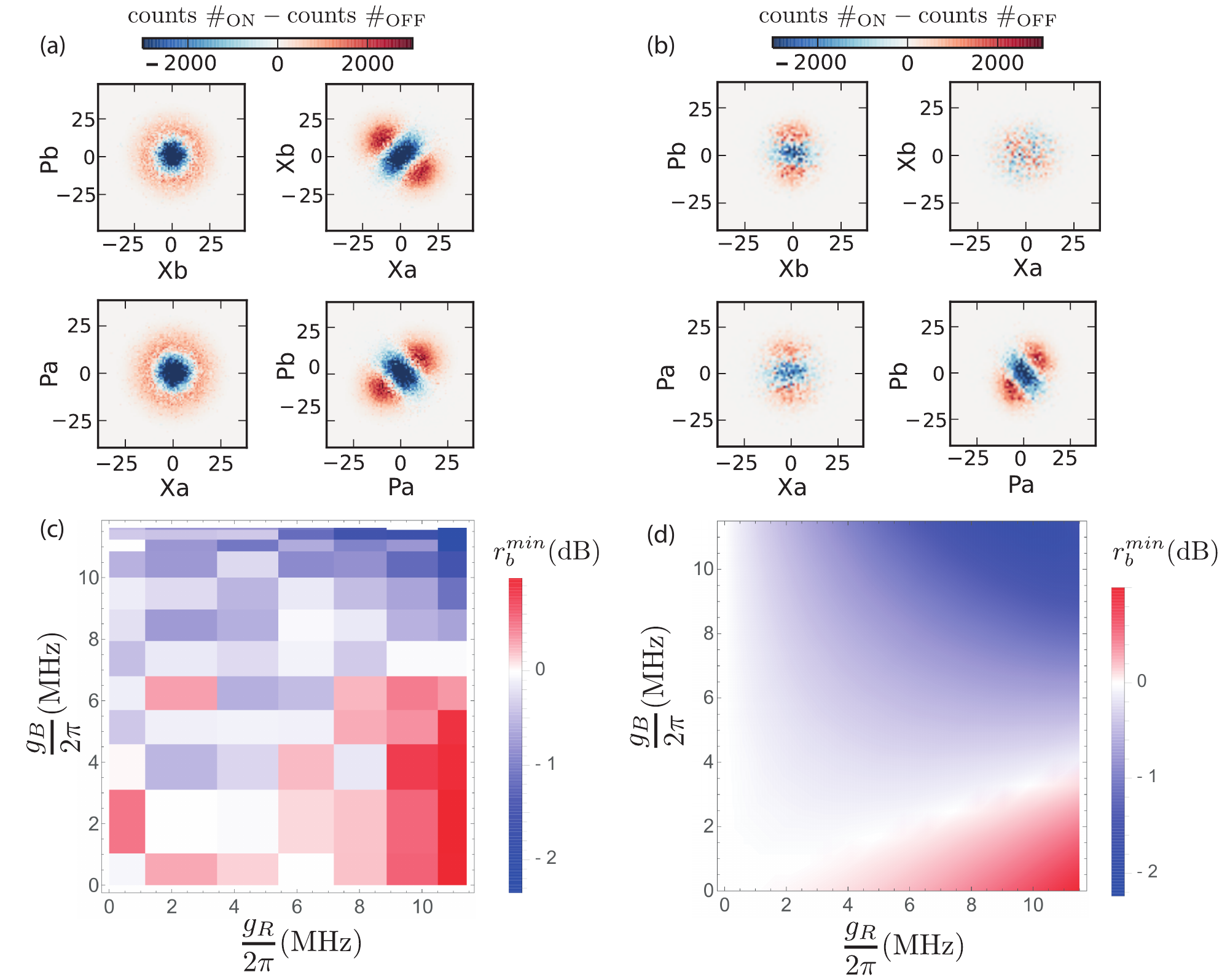}
\caption{(a) Result of the subtraction between quadrature histograms (in number of counts per bins of $1\times 1$) with the pump turned on and off. The red pump amplitude is here null and there is no detuning $\omega_\mathrm{eff}=0$.With an independent determination of the gain of the device (here, $G=16~\mathrm{dB}$), these figures can be used as a calibration of the quadrature axes, here in square root of photon number in the modes. (b) Similar measurement in the case where $\omega_\mathrm{eff}=2\pi\times 26~\mathrm{MHz}$ and $g_B=g_R=2\pi\times 12.6~\mathrm{MHz}$. (c) Measured and (d) calculated squeezing parameter $r_b$ for the $b$ mode as a function of the ``blue'' and ``red'' pump coupling rates $g_B$ and $g_R$. The measured $r_b$ is inferred from the distributions in the plane $X_b,P_b$.}
\label{squeezing}
\end{figure}

In the frame rotating at $\omega_{a,b} + \omega_{\textrm{eff}}$, these frequencies correspond to $\pm \omega_{\textrm{eff}}$, i.e. the positive and negative images of the frequency $\omega_{\textrm{eff}}$ of each effective mode $c$ and $d$. The coupling rate of the ``blue'' pump $g_B$ is determined by matching the average emission rate $\langle a_{out}^\dagger[\omega] a_{out}[\omega] \rangle$ that can be calculated using the input-output theory to the difference between the power spectral densities measured with the pump turned on or off. In the following measurement, we set the pump amplitude and frequency such that $g_B = 2 \pi \times 12.6$~MHz and $\omega_{\textrm{eff}} = 2 \pi \times 26$~MHz. The latter is chosen to reach the third regime in Fig.~\ref{spectro}a with limited pump power.

On top of the ``blue'' pump, we then apply the ``red'' pump at an amplitude $\mathcal{A}_R$. The measured power spectral density reveals a hybridization of the effective modes in the simulated ground state. Indeed, each of the spectral peaks at $\pm \omega_{\textrm{eff}}$ splits into two peaks separated by the coupling rate $g_R \propto \mathcal{A}_R$ (see Figs.~\ref{spectro}b,d), leading to a total of four frequency peaks per measured output. Such a splitting is analogous to the vacuum Rabi splitting one observes for the quantum Rabi model. Our quantum simulation allows to transfer this property to the outgoing cavity fields. As seen on Figs.~\ref{spectro}b,d, the two middle peaks get closer as the coupling $g_R$ increases until they eventually merge when $g_R \approx g_B$. At this particular point, the frequency of the hybrid effective mode thus collapses to zero. Such a collapse can have important physical consequences, as in the Dicke model where it is associated to a quantum phase transition~\cite{Emary2003}. Input-output theory allows us to qualitatively reproduce the measured spectral density features (Figs.~\ref{spectro}c,e) and provides a calibration of the coupling rate $g_R$ as a function of the ``red'' pump amplitude $\mathcal{A}_R$. This calibration is shown as a scale of $g_R$ on the right axis of Figs.~\ref{spectro}c,e.

As we have shown in Ref.~\cite{Fedortchenko2017}, the spectral peaks indicate frequencies for which one quadrature of the radiated field is squeezed while the other one is anti-squeezed. Maximal squeezing is expected for $g_R = g_B\approx \omega_\mathrm{eff}/2$, since for this value the ultrastrong coupling condition is well established. Furthermore, the ground state entanglement between the effective modes $c$ and $d$ that results from ultrastrong coupling here corresponds to the two-mode squeezing of the fields radiated from modes $a$ and $b$, similarly to the EPR state created in the amplification regime with only vacuum fluctuations at the input of the modes~\cite{Flurin2012a}.

We characterize squeezing by measuring the distribution of field quadratures for both output modes: $\hat{X}_a = \frac{\hat{a}_\mathrm{out} + \hat{a}_\mathrm{out}^\dagger}{2}$ and $\hat{P}_a = \frac{\hat{a}_\mathrm{out} - \hat{a}_\mathrm{out}^\dagger}{2i}$ for the output of mode $a$ and similarly defined $\hat{X}_b$ and $\hat{P}_b$ for the output of mode $b$. The heterodyne signal is amplified, down-converted to below 100~MHz and digitized using an acquisition board. We interleave the measurement $10^6$ times with the pumps being alternatively turned on and off in order to remove the contribution of the potentially drifting added noise~\cite{Eichler2011a, Eichler2012, Menzel2010, Flurin2015}. We characterize our measurement scheme by first focusing on the known case where the blue pump is not detuned $(\omega_\mathrm{eff}=0)$ and the red pump is turned off $(g_R=0)$ so that an EPR state is generated~\cite{Flurin2015}. On Fig.~\ref{squeezing}a are shown the result of the subtraction of the measured mode quadrature distribution corresponding to the pump turned off to the distribution when the pump is turned on. The single-mode quadrature distributions (left panels) are uniformly distributed in phase, which is expected since each mode $a$ and $b$ is occupied by a thermal state. In contrast, the cross quadrature distributions show a clear correlation between the quadratures of modes $a$ and $b$ (right panels of Fig.~\ref{squeezing}a), which is directly linked to the amount of entanglement in the EPR state (here 9~e-bits of logarithmic negativity for a gain of 16~dB)~\cite{Flurin2015}. 

The distributions of quadratures change drastically in the ultrastrong coupling regime when $g_B=g_R= 2\pi\times 12.3~\mathrm{MHz}$ (Fig.~\ref{squeezing}b). Indeed, while there are still two-mode correlations (bottom right panel), the single-mode distributions also show evidence of squeezing (left panels). It is consistent with our claim in Ref.~\cite{Fedortchenko2017} that both single-mode and two-mode squeezing arise in the ultrastrong coupling regime. We reconstruct the covariance matrix $\mathcal{V}_{ij} = \langle x_i x_j \rangle - \langle x_i \rangle \langle x_j \rangle$, where $\textbf{x} = \left\{ \hat{X}_a, \hat{P}_a, \hat{X}_b, \hat{P}_b \right\}$. It can be block-diagonalized to find the eigenvalues $\sigma_a^{min, max}$ and $\sigma_b^{min, max}$ of the single-mode covariance matrices for the modes $a$ and  $b$. These eigenvalues correspond to variances of the maximally squeezed and anti-squeezed quadratures of the propagating modes $a_{out}$ and $b_{out}$. Squeezing is quantified using a squeezing parameter $r_{a,b} = 10\log_{10} \left(\frac{\sigma_{a,b}^\mathrm{min,on}-\sigma_{a,b}^\mathrm{off}}{G_{a,b}\sigma_{vac}}+1\right)$, where $G_{a,b}\sigma_{vac}$ is the variance of the vacuum fluctuations once amplified by the detection setup, and is calibrated using the independently known covariance matrix of the EPR state (see Fig.~\ref{squeezing}a)~\cite{flurin:tel-01241123}. We have measured the single-mode squeezing parameter for many values of the coupling rates $g_B$ and $g_R$ (Fig.~\ref{squeezing}c). As expected, they reach a minimum for $g_B = g_R \simeq \frac{\omega_{\textrm{eff}}}{2}$. Note that maximum squeezing is not the same for the two modes, $r_a^{min} = -1$ dB while $r_b^{min} = -2.3$ dB. This is consistent with asymmetric output coupling rates $\kappa_a \neq \kappa_b$. Interestingly, when the red pump amplitude decreases with the blue one remaining constant, the single-mode squeezing parameter increases and even becomes positive. This behavior is reminiscent of the EPR state (blue pump only at zero detuning) for which a thermal state establishes in each mode. Predictions based on input-output formalism for the single-mode squeezing parameter reproduce the measurements qualitatively~\cite{Fedortchenko2017}.

\begin{figure}
\includegraphics[scale=0.43]{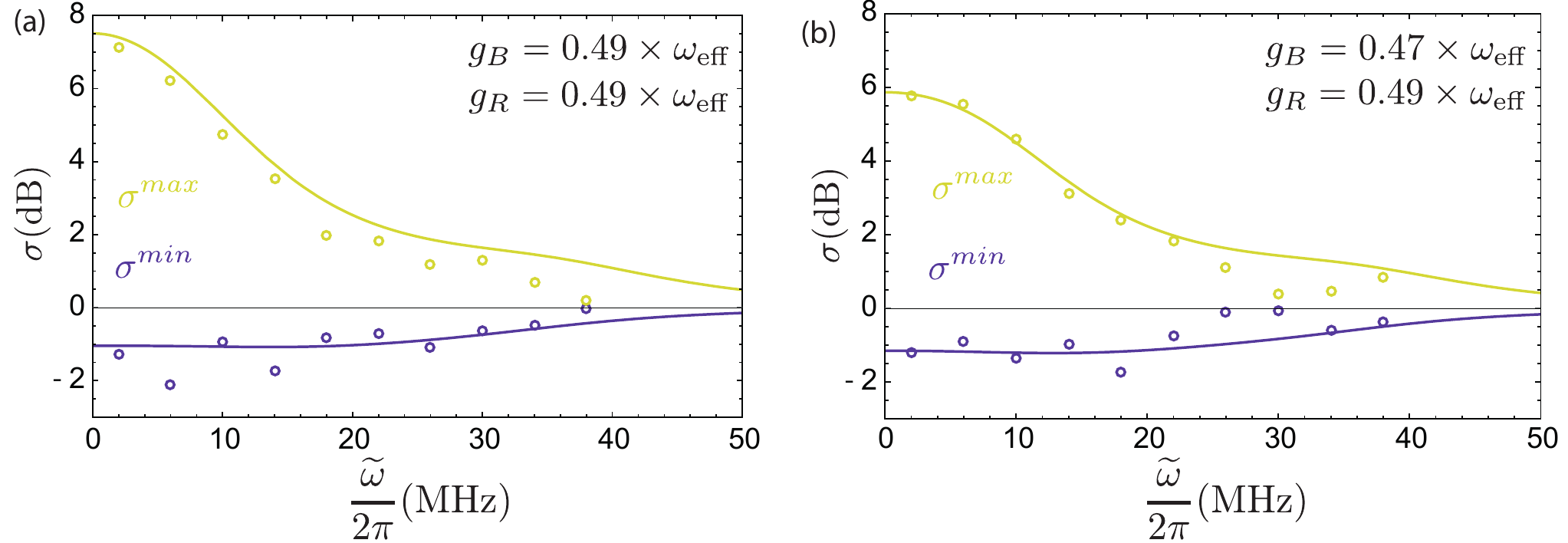}
\caption{Measured minimum and maximum variance $\sigma_a^\mathrm{min}$ and $\sigma_a^\mathrm{max}$ of a single quadrature of mode $a$ as a function of the detuning $\delta \omega$ referred to the rotating frame (corresponding to $\omega_a + \omega_{\textrm{eff}} + \widetilde{\omega}$ in the laboratory frame). The variances in dB are referred to the measured variance when the pumps are turned off. Statistical errors bars are smaller than the dot size but systematic errors seem to remain of the order of 1~dB at worst. Figures (a) and (b) correspond to two different coupling rates $g_B$ of the blue pump (see labels on top). Full lines correspond to the expectation value calculated using input-output theory.}
\label{qrs}
\end{figure}

We also characterize the two-mode squeezing by considering the variance
of collective variables $\hat{X}_a-\hat{X}_b$, $\hat{X}_a+\hat{X}_b$, $\hat{P}_a-\hat{P}_b$ and $\hat{P}_a+\hat{P}_b$. The determination of a squeezing parameter is highly sensitive to the amplification factors of the measurement lines $G_a = (4.2 \pm 0.9) \times 10^{-8}$ V$^2$  and $G_b = (9.2 \pm 0.9)\times 10^{-8}$ V$^2$. 
The uncertainty on their exact values at high pump amplitudes thus leads to a large uncertainty on the squeezing parameter and we could only demonstrate that the variance of $\hat{X}_a-\hat{X}_b$ and $\hat{P}_a+\hat{P}_b$ is smaller than the vacuum fluctuations, hence demonstrating the presence of two-mode squeezing in the regime of effective ultrastrong coupling.

Finally, we have measured the variances of the two quadratures of the modes that show maximal and minimal variance as a function of frequency and for two values of the blue pump amplitude in a regime where $g_B < \omega_{\textrm{eff}}/2$ (Fig.~\ref{qrs}). The field is squeezed below  vacuum fluctuations $(\sigma^\mathrm{min} < 0~\mathrm{dB})$ over a range of frequencies comparable to $2 \omega_{\textrm{eff}}$. Besides, a characteristic inflexion can be observed over the same bandwidth for the antisqueezing component $\sigma^\mathrm{max}$. These features are quantitatively reproduced by our model (lines).

In conclusion, we have realized an analog quantum simulation of two ultrastrongly coupled harmonic oscillators using a Josephson mixer. We have demonstrated spectroscopic evidence of mode hybridization and mode collapse in the USC ground state. We have also detected simultaneous single-mode and two-mode squeezing of the emitted field, which is related to the entangled nature of the ground state in the ultrastrong coupling regime~\cite{Fedortchenko2017}.  Finally, we have measured the single-mode squeezing and antisqueezing as a function of frequency separately for each field quadrature and have observed vacuum squeezing over the whole bandwidth of the effective mode.

\begin{acknowledgements} Nanofabrication has been made within the consortium Salle Blanche Paris Centre. This work was supported by the EMERGENCES grant QUMOTEL of Ville de Paris, by the French Agence Nationale
de la Recherche (GEARED project No. ANR-14-CE26-0018, SemiQuantRoom project No. ANR14-CE26-0029)
and by the PRESTIGE program, under
the Marie Curie Actions-COFUND of the FP7. The authors acknowledge F. Portier, A. Keller and G. Steele for interesting discussions.
\end{acknowledgements}

\end{document}